\documentclass[namedreferences]{solarphysics}
\usepackage[optionalrh]{spr-sola-addons} 
\usepackage{epsfig}          
\usepackage{graphicx}        
\usepackage{amssymb}        
\usepackage{color}           
\usepackage{url}             

\begin{document}

\begin{article}

\begin{opening}

\title{The Relation between the Radial Temperature Profile in the Chromosphere
and the Solar Spectrum at Centimeter, Millimeter,
  Sub-millimeter, and Infrared Wavelengths}

\author{V.~\surname{De la Luz}$^{1}$\sep
        M.~\surname{Chavez}$^{1}$\sep
 E.~\surname{Bertone}$^{1}$\sep
        G.~\surname{Gimenez de Castro}$^{2,3}$
       }
\runningauthor{De la Luz et al.}
\runningtitle{The Solar Temperature Profile}

   \institute{$^{1}$ Instituto Nacional de Astrofisica, Optica y Electronica,
  Tonantzintla, Puebla, Mexico, Apdo. Postal 51 y 216, 72000
                     email: \url{vdelaluz@inaoep.mx} email: \url{mchavez@inaoep.mx} email: \url{ebertone@inaoep.mx}\\
              $^{2}$ Centro de R\'adio Astronomia e Astrof\'{\i}sica Mackenzie, UPM,
                 R. da Consola\c{c}\~ao 896, 01307-902, S\~ao Paulo, SP, Brazil
                     email: \url{guigue@craam.mackenzie.br} \\
              $^{3}$ Instituto de Astronom\'{\i}a y F\'{\i}sica del Espacio, Ciudad Universitaria,
                 CC 25, 1428, Buenos Aires, Argentina\\
             }

\begin{abstract}
Solar observations from millimeter to ultraviolet wavelengths show 
that there is a temperature minimum between photosphere and chromosphere.
Analysis based on semi-empirical
models locate this point at about 500~km 
over the photosphere. The consistency of these models
has been tested by means of millimeter to infrared
observations.

In the present work, we show that variations of the theoretical radial
temperature  profile near the temperature minimum impacts the brightness temperature at centimeter, submillimeter, and infrared 
wavelengths, but the millimeter wavelength emission remains unchanged. We found a region between 500 and 1000 km over the photosphere that remains
hidden to observations at the frequencies under study in this work.

\end{abstract}
\end{opening}

\section{Introduction}
Compared to the solar radius, the chromosphere is a thin region in the upper solar atmosphere 
where, as a function of formation height, spectral lines (e.g., Ca~{\sc ii} and Mg~{\sc ii}) change from absorption to emission
 \cite{1961psc..book.....A}. Additionally, it has been
found that molecular CO bands (between 2.1 and 15.5~$\mu$m) and infrared, millimeter, and sub-millimeter continuum emission are
produced in that region \cite{1957IAUS....4..263H,1978ApJ...225..665A,2000ApJ...536..481U,2003ASPC..286..419A}.
The standard interpretation of the solar radio emission (from meters to sub-millimeter waves) states that the main contributor to the emission is the Bremsstrahlung process in Local Thermodynamic Equilibrium (LTE) and predicts that the observed effective height of flux emission formation depends on frequency \cite{2011SoPh..273..309S}.  

\begin{figure}
   \centerline{
\includegraphics[height=0.8\textwidth]{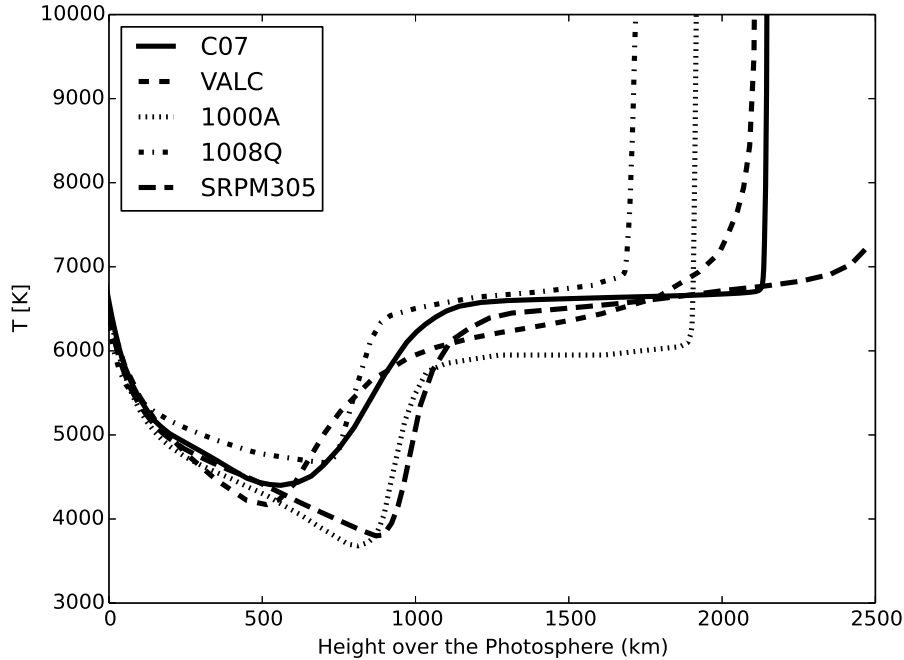}}
\caption{The Temperature Minimum region between 100 and 1100 km over the photosphere of the C07 (continuos line), VALC (dashed line),  1000 (small dashed line), 1000A (dotted line), 1008Q (dot-dashe line) and SRPM305 (large dashed line) models.}
\label{minimumTemperatures2.eps}
\end{figure}

In general, semi-empirical models assume a one-dimensional approach to the physical conditions of the chromosphere. This category includes the models VALC from \inlinecite{1981ApJS...45..635V}, SRPM305 from \inlinecite{2006ApJ...639..441F}, C07 from \inlinecite{2008ApJS..175..229A}, and the cold [1000A], and hot [1008Q] models from \inlinecite{2011JGRD..11620108F}.
These models provide the best theoretical fit so far achieved to the observed spectra and represent a good approximation because the horizontal scale is
larger than the vertical scale of the elements under study \cite{2006ApJ...639..441F}. All models share a common feature: a region 
where the atmosphere reaches
a temperature minimum, 
where the temperature gradient is inverted
and the radial temperature increases gradually outward until the so-called Transition Region
(Figure \ref{minimumTemperatures2.eps}). These 1D semi-empirical models show that radiation at low frequencies emerges from high atmospheric layers \cite{2008ApJS..175..229A},
indicating a close correlation between the observed brightness temperature ($T_{b}$) and the radial temperature profiles. Such a correlation has been observed
not only in the Sun, but also in other stars \cite{2013A&A...549L...7L}. Whilst the mentioned models provide the best representation of the high
atmospheric layers, there are still significant inconsistencies  between observed brightness temperature ($T_b$) and synthetic spectra at millimeter-IR regimes \cite{1981ApJS...45..635V,2008ApJS..175..229A,2011ApJ...737....1D},
in particular at $\nu < 30$~GHz, and at sub-millimeter and infrared frequencies, in the interval from 400 to 3000~GHz.
Since the region of temperature minimum has long been thought of as the location where infrared to millimeter emission originates, we consider convenient to
explore the effects of modifying the temperature structure of several semi-empirical models on the spectrum at this wavelength in this regime.

In this work, we improve the code PakalMPI \cite{2010ApJS..188..437D} to compute the physical conditions (pressure, temperature, and density) in hydrostatic equilibrium of the solar chromosphere. 
We compute the absorption and emission processes from  millimeter to infrared wavelengths in the solar chromosphere using the
C07 model (we have used the C07 model as a reference because its temperature structure represents an approximate average of the rest of the models) with two modifications, 
in which we artificially modify the temperature, using as threshold the minimum and maximum temperature reported in the
models C07, VALC, SRPM305, 1000A, and 1008Q. We add 270 K and substract 720 K in the
radial temperature profile around the temperature minimum.

The modifications are aimed at testing the diagnostic capabilities of the
centimeter-infrared emission, using the Chromospheric Solar Millimeter-wave Cavity or CSMC \cite{2013ApJ...762...84D} to 
trace the properties of the temperature minimum region
and at establishing the role of the morphology of the temperature minimum 
in the discrepancy between the observed and the synthetic spectrum at the frequencies under study. 
In section \ref{model} we define the model under study and the modifications in the temperature minimum region; 
in section \ref{computations} we show the convergence test made to our code and we
explain the numerical computations of the physical parameters of the solar atmosphere; 
in section \ref{results} we present the results and finally, in section \ref{conclusions}, the conclusions and remarks are given.

\section{The Models}\label{model}

As a representative case, we have choosen the C07 model because its temperature
minimum (4400 K) is between the range of values reported by the VALC (4170 K), SRPM305 (3800 K), 1000A (3680 K), and 1008Q (4670 K) models. 

We modified the temperature profiles adding to the C07 model a Gaussian function, with amplitudes
of +270 K (this model is called C07+270K) and -720 K (C07-720K), a half width at half maximum of 522 km,
and centered close the temperature minimum at a height of 560 km over the solar photosphere.

\begin{figure}
   \centerline{
\includegraphics[width=1.0\textwidth]{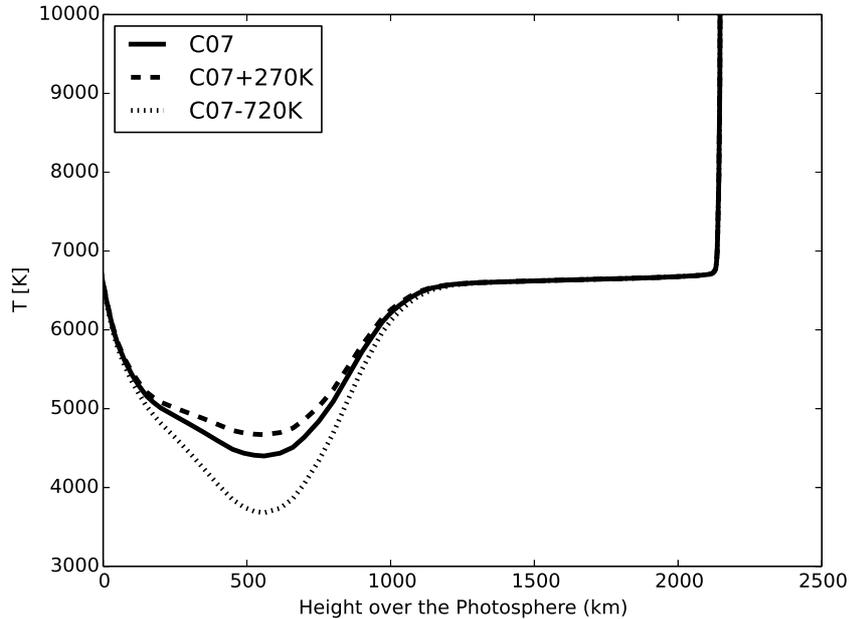}}
\caption{Solar radial temperature profiles for the original C07 model (continuos line), C07+270K (dashed line) and C07-720K 
(dotted line).}
\label{temperature.png}
\end{figure}

Note that the height selection corresponds to the region commonly associated with the
layer where the mm to IR spectral properties are sculpted (see Figure \ref{temperature.png}).


\section{Computations}\label{computations}
We improved PakalMPI \cite{2010ApJS..188..437D} in order to include the computations of density and pressure in a stratified atmosphere.
We followed the mathematical theory by \opencite{1973ApJ...184..605V} using our approximation for the hydrogen departure coefficient
 (the $b_1$ parameter) by \inlinecite{2011ApJ...737....1D} taking into account 20 ionic species at different
ionization stages. 
The code uses the radial temperature profile as a reference
and modifies the density profile to stabilize the pressure in order to ensure the hydrostatic equilibrium. 
As the
variables are inter dependent, 
several iterations are needed until the equilibrium is reached,
starting from a set of
initial values close to the possible
solution. We found that 
the initial C07 profiles of density and pressure profiles convergence to
an atmosphere in hydrostatic equilibrium
for  both the C07+270K and C07-720K models.

\begin{figure}
   \centerline{
\includegraphics[width=1.0\textwidth]{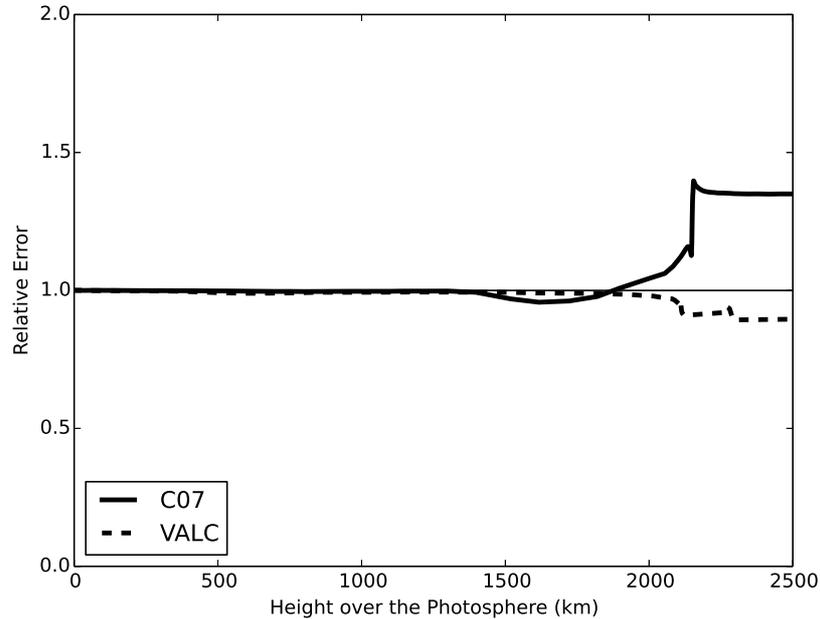}}
\caption{The relative error between the input model and the output model for the C07 (continuos line) and the VALC (dashed line) models. The results shows that for the first 2150 km, the average relative error for C07 model is around 2.6\%. For the case of the VALC model, the average relative error is of 2.0\% between.
However after 2150 km the relative error grows at 36\% for the case of the C7 model and 8\% for the VALC model.
}
\label{relative_error}
\end{figure}

To test the numerical stability of our code, we performed the following computations: we used a pre-equilibrated model as input model (temperature, density and pressure), then the result is compared (divided) by the input model.
Since we are using an approximation for the $b_1$ parameter from VALC model, we tested our code with both C07 and VALC. 
In the case of VALC model the average relative error for the density 
is around 4\%. 
However, the relative error of the C07 model is almost
zero up to a height of 2150 km over the photosphere, then the relative error increases to 36\% (Figure \ref{relative_error}).  
The main differences in density above 2150 km are most probably due to our approximation of the $b_1$ parameter (see details in \opencite{2011ApJ...737....1D}). For the case of VALC, the errors are low (as expected) because we are using the same $b_1$ values, but for the case of the C07 model the differences in the dynamic space (temperature and density) with respect to VALC produce an under abundance of gas. We, nevertheless, would like to remark that the analysis presented in this work is focused at altitudes lowers than 2150 km, therefore the results are not affected by this inconsistency at the highest layers.

We also used the code {\tt PakalMPI} to solve the radiative
transfer equation and compute the emission and absorption processes for each model between 2~GHz ($\lambda=15$~cm) and 10~THz ($\lambda=3$~$\mu$m). The height $h=0$~km corresponds to the photosphere and
increases outward. In all the computations, three opacity functions are included: Classical Bremsstrahlung
\cite{1979ApJS...40....1K}, H${}^-$ \cite{1996ASSL..204.....Z}, and
Inverse Bremsstrahlung \cite{1980ZhTFi..50R1847G}. 

We also calculated the local emission efficiency $E_l$
defined as
\begin{equation}\label{efiloca}
E_l = 1-\exp{(-\tau_{\rm local})} \ ,
\end{equation}
where $\tau_{\rm local}$ is the local optical depth. The local emission efficiency ranges between 0 (optically thin) and 1 (optically thick). In the
first case ($E_l \sim 0$), the atmosphere is not efficient in generating or absorbing
radiation while in the second case ($E_l \approx 1$) the atmosphere may be approximated by
a black body at the frequencies under study (for details, see \opencite{2013ApJ...762...84D}). Finally,
we computed the brightness temperature $T_b$ in the same spectral range.

\section{Results}\label{results}

\begin{figure}
   \centerline{
\includegraphics[width=1.0\textwidth]{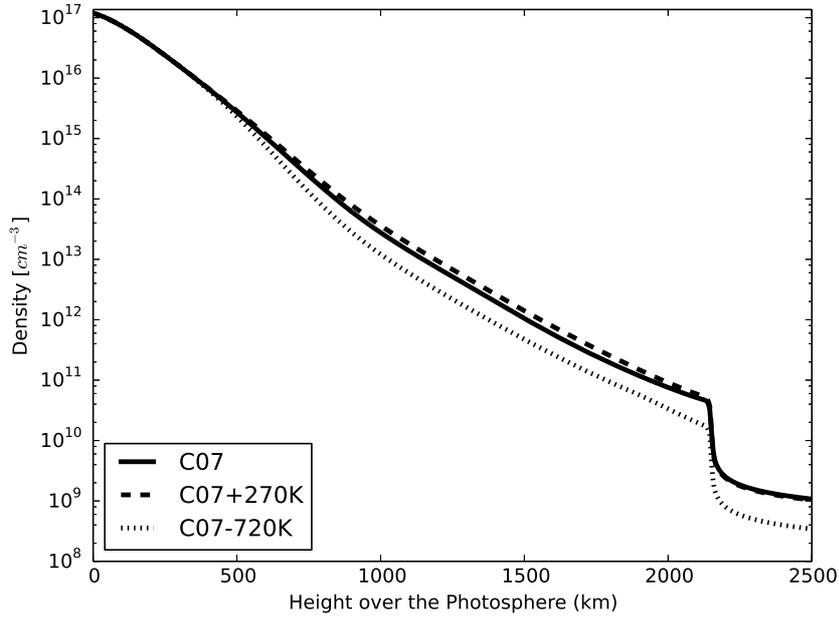}}
\caption{The number density for the original C07 model (continuos line), C07+270K (dashed line) and C07-720K (dotted line).}
\label{hydrogen.eps}
\end{figure}

\begin{figure}
   \centerline{
\includegraphics[width=1.0\textwidth]{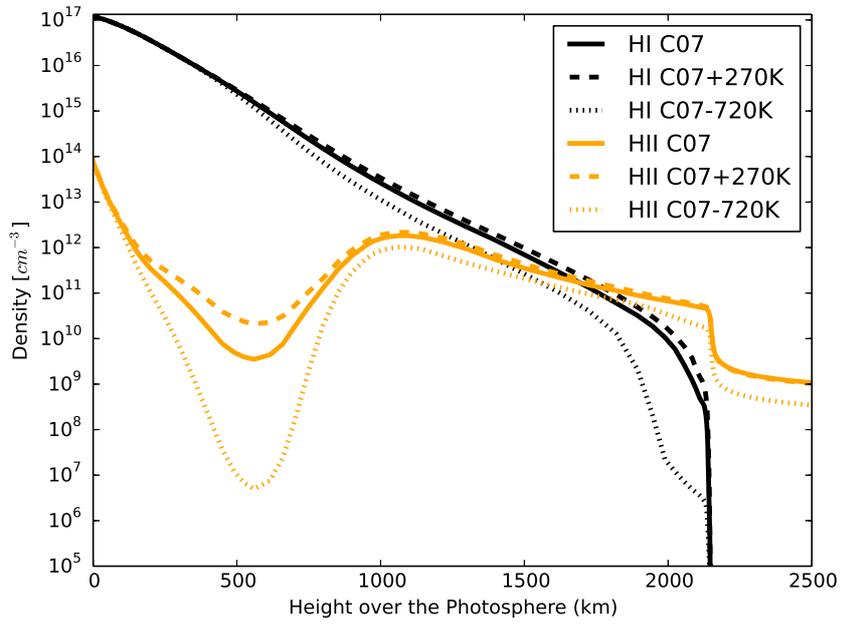}}
\caption{The neutral Hydrogen (HI, black color lines) and proton (HII, yellow/gray color lines) number density for the original C07 model (continuos line), C07+270K (dashed line) and C07-720K (dotted line).}
\label{HI.eps}
\end{figure}

\begin{figure}
   \centerline{
\includegraphics[width=1.0\textwidth]{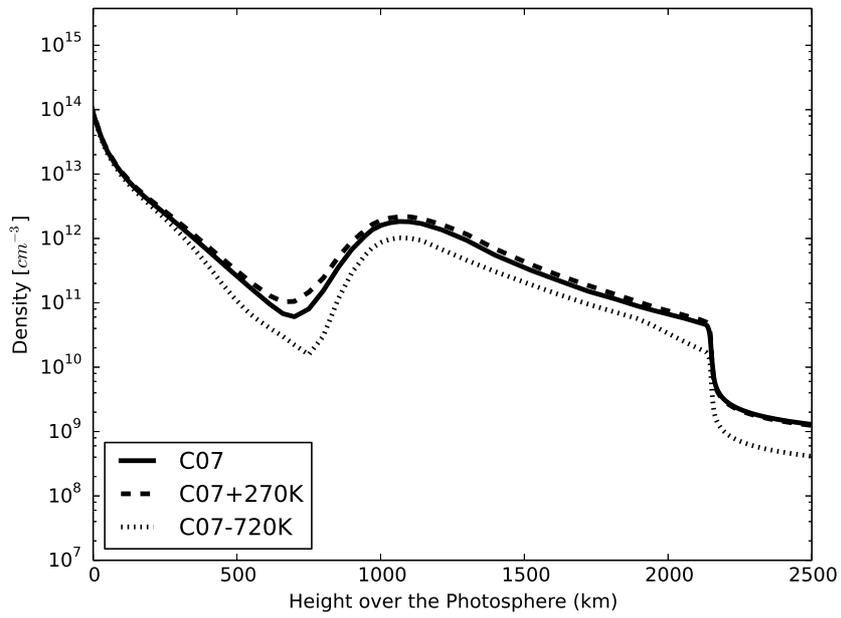}}
\caption{The electronic density for the original C07 model (continuos line), C07+270K (dashed line) and C07-720K (dotted line).}
\label{elecrons.eps}
\end{figure}

We analyzed the most important physical quantities related with the emission/absorption processes: the hydrogen density and 
its ionization state
the electronic density, the local opacity, the optical depth and the brightness temperature. The global behavior of the chromospheric models in terms of the altitude are ilustrated in Figures \ref{hydrogen.eps} to \ref{elecrons.eps}. 

Figure \ref{hydrogen.eps} shows the variation of the total gas density with altitude for the three models; C07, C07+270K, and C07-720K. 
We found that along the first 500 km the models are very similar. However, above this height, a dependence with temperature minimum region is observed, such that the density decreases for lower values of the temperature minimum region (C07-720K) or increase at higher temperatures (C07+270K) with respect to the C07 model. In Figure \ref{HI.eps} we show individual neutral hydrogen (HI) and  proton (HII) densities for the three models.
Note that, at about 2200 km over the photosphere, hydrogen becomes 
fully ionized, as a consequence of the temperature increment towards the transition region. 
In the region between 100 and 1000 km over the photospere a decrement in the proton density is observed, 
the lower temperature model (C07 - 720K) being the most
affected.
The electronic density is plotted in Figure \ref{elecrons.eps}, that shows a characteristic region of high density,
between 800 and 2200 km, and a decrement in the profile between 200 and 1000 km over the photosphere. 
In this region, 
free electrons in our approximation are produced by ionization of hydrogen as well as 19 other ionic species.
Pakal calculates the electronic densities allowing LTE departure for hydrogen only. The rest of the species is considered in LTE.

\begin{figure}
   \centerline{
\includegraphics[width=0.8\textwidth]{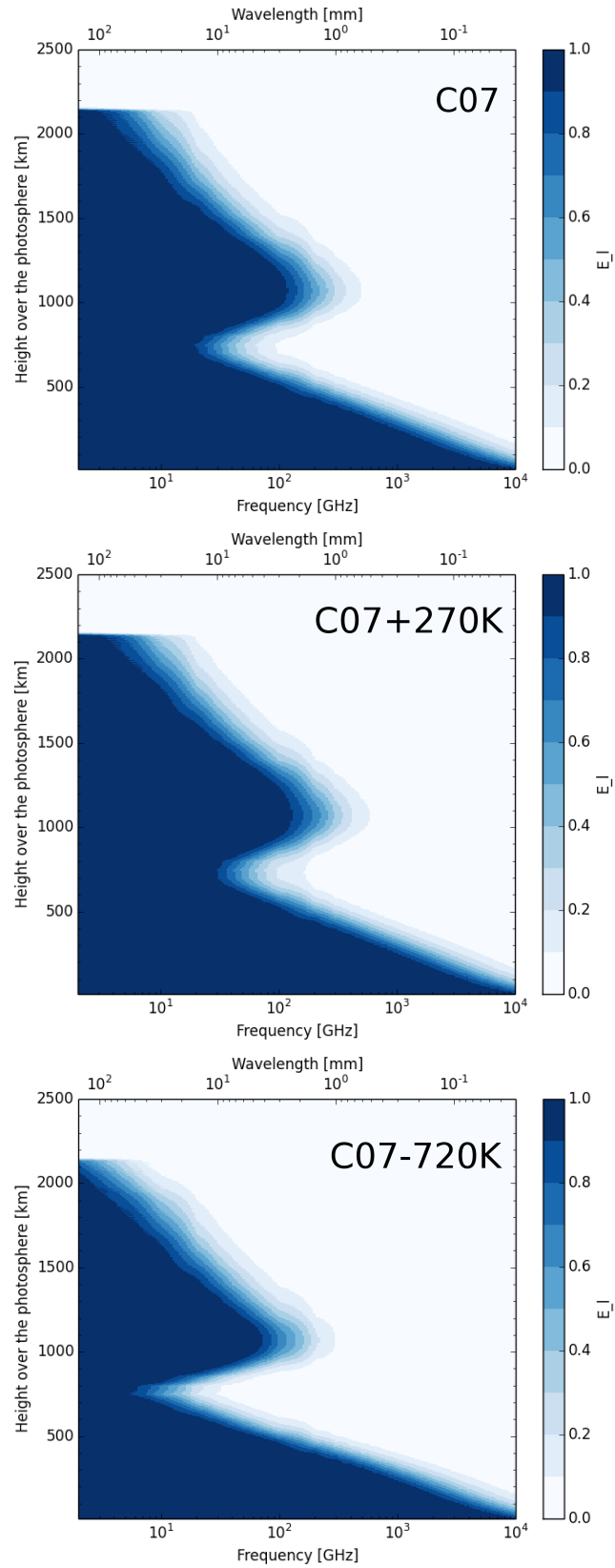}}
\caption{The CSMC for the C07 (top), C07+270 (middle), and C07-720K (bottom) models. The color scale on the right of each panel indicate the value of the local emissivity.}
\label{C07.eps}
\end{figure}

Following the analysis presented in \inlinecite{2013ApJ...762...84D}, we found that the CSMC is present in the C07, C07+270K, and C07-720K models as displayed
in the upper, middle and bottom panels of figure \ref{C07.eps}, respectively. 
It is interesting to note that for lower values of
the temperature minimum the CSMC appears at lower frequencies, and the second optically thick region (peak) of the CSMC is present in the three models.

In order to analyze the tomography properties of the radio-infrared interval, we 
computed the height where the atmosphere becomes optically thick ($\tau(\nu) = 1$), which provides 
an estimate of the altitude where the emission originates.

These results are illustrated in Figure \ref{nuvsheightwt1.eps}, where we show that the region 
between 500 and 1000 km over the photosphere (gray rectangle on the plot) remains hidden to observations at radio-infrared wavelengths.

\begin{figure}
   \centerline{
\includegraphics[width=1.0\textwidth]{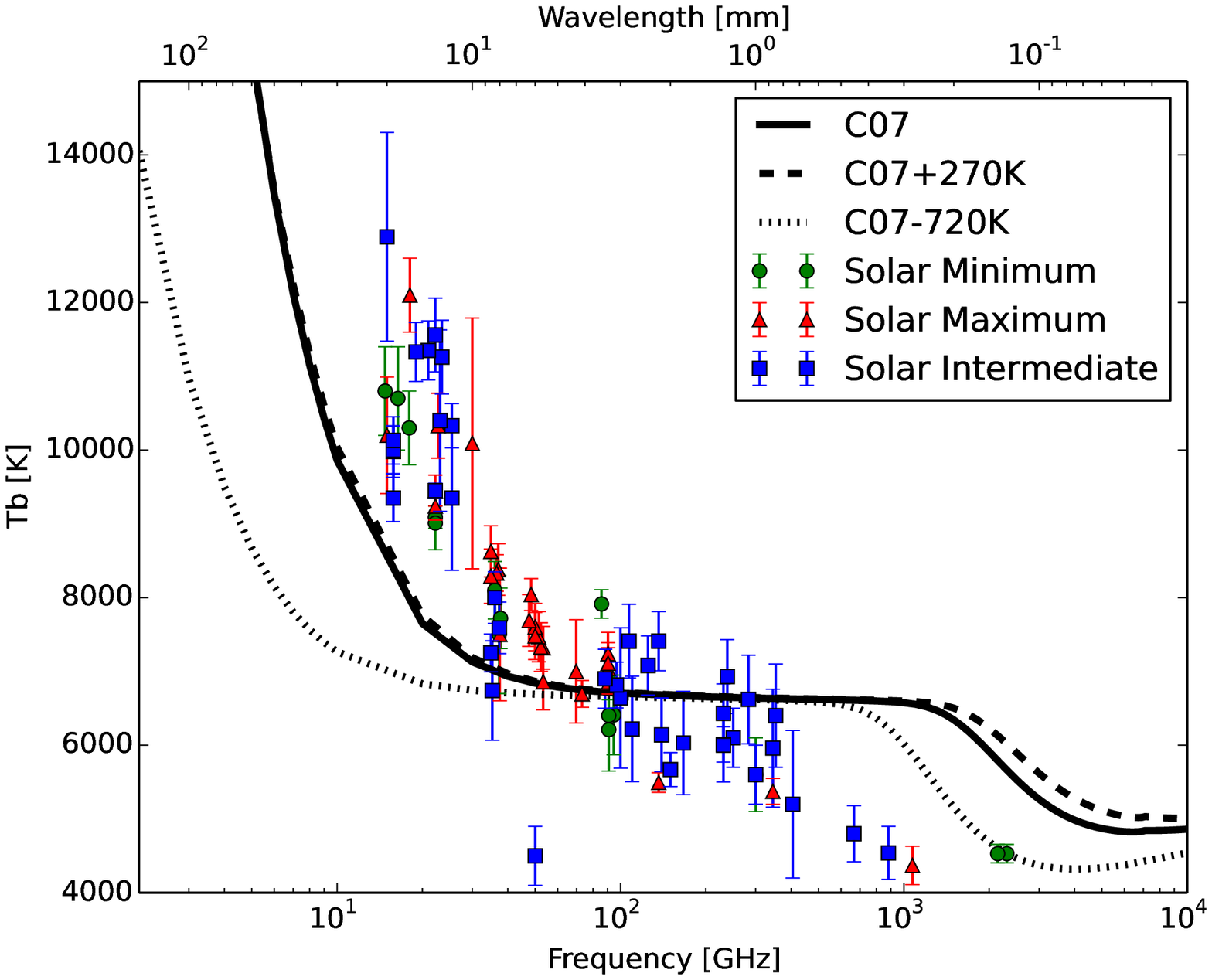}}
\caption{Solar synthetic spectrum for the original C07 model (continuos line), C07+270K (dashed line) and C07-720K (dotted line)
and solar observations for three different stages of solar activiy. We found that modifications in the
minimum temperature regions affect the radio and infrared regions of the spectrum.}
\label{spectrum.eps} 
\end{figure}

Figure \ref{spectrum.eps} shows the synthetic spectra for the three models together with the observed data collected from
\inlinecite{2004A&A...419..747L}. We show that the modifications in the
region of the temperature minimum impact the radio, submillimetr, and infrared frequencies, while 
the millimeter
region remains unchanged.

\section{Conclusions}\label{conclusions}

We know that Bremsstrahlung and the H${}^-$ opacity dominate the absorption of the solar chromosphere at radio - infrared frequencies \cite{2011ApJ...737....1D}.
At altitudes greater than 500 km, Bremsstrahlung is the major contributor of the total opacity.
However, at about this altitude, the temperature drops under the ionization temperature threshold for hydrogen atom. This 
results in a low number of ions and electrons that finally produce a locally thin atmosphere (the CSMC).
Outward, the temperature rises again and at 6000 K the atmosphere becomes again locally optically thick, creating a second region of emission (the peak of the
CSMC) at altitudes between 800 and 1500 km over the photosphere.

\begin{figure}
   \centerline{
\includegraphics[width=1.0\textwidth]{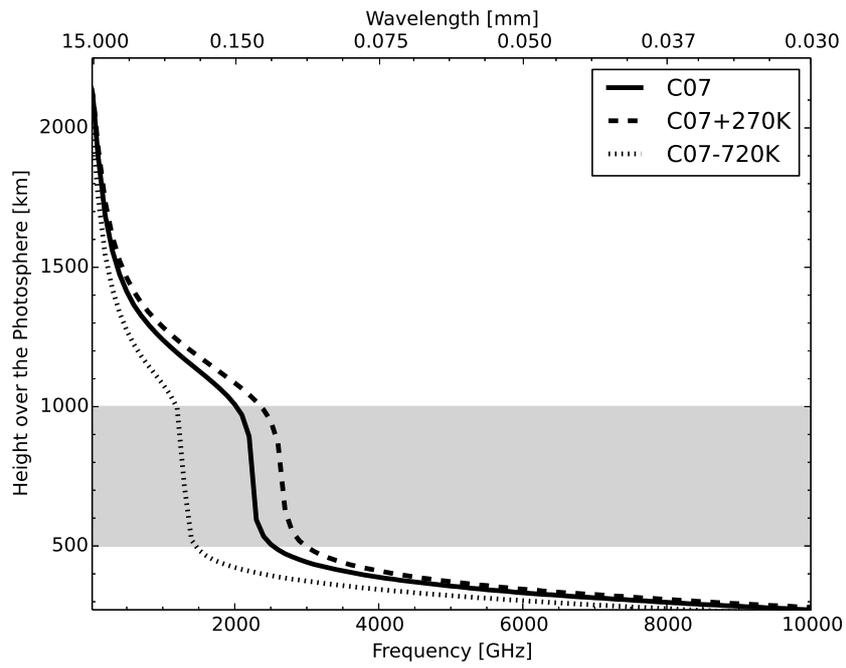}}
\caption{Optical depth for the original C07 model (continuos line), C07+270K (dashed line) and C07-720K (dotted line). We found
that between 500 and 1000 km over the photosphere the atmosphere remains hidden for the radio-infrared observations.}
\label{nuvsheightwt1.eps}
\end{figure}

The peak of the CSMC is responsible of the drop of the optical depth at about 500 km as depicted in Figure \ref{nuvsheightwt1.eps}. 
This decrease creates a hidden region for the observations at radio-infrared frequencies covering altitudes from 500 to 1000 km over the photosphere. 


In spite of not being directly accessible to observations, the impact of the temperature minimum morphology can be ascertained at other adjacent spectral windows. We have seen that, while $T_b$ does not change even under significant modifications of the depth of the temperature minimum, the effects of these modifications are 
noteworthy
at centimeter, sub-millimeter and infrared bands. The changes in $T_b$ can be ascribed to two main
reasons:
%
for the case of the infrared region, the radial temperature is the responsible of the change of their H${}^-$ opacity while the gas density
is the responsible of the change of the Bremsstrahlung opacity at centimeter and submillimeter frequencies. We found that 
in order to reduce the $T_b$ discrepancies between models and observations, that are evident in the frequency interval
400~GHz--3~THz (submillimeter region), the local Bremsstrahlung opacity in the peak of emission of the CSMC 
should be reduced, not the temperature minimum value. However, a decrease in temperature causes a decrease in density at higher altitudes and also results in the
decrease of the brightness temperature at centimeter wavelengths.

According to the behavior depicted in Figure \ref{nuvsheightwt1.eps}, the morphology of the temperature minimum cannot reduce the discrepancies between observations and theory. To this aim, a variety of solutions or combinations of them can be 
proposed, all of which require
a decrease of the frequency of the CSCM peak:
(i) increasing the radial temperature profile at these altitudes, (ii) including other
processes such as the Farley-Buneman instability \cite{2008A&A...480..839F} that reduce the electronic density or the number
of ions at these altitudes, (iii) changing the geometric symmetry 
as the plain parallel approximation might not be correct at the location
of the peak of the CSMC, (iv) include magnetic structures as suggested by \cite{2007ApJ...664.1214P,2010SoPh..261...53V}.


\begin{acks}
CGGC is grateful to FAPESP (Proc. 2009/18386-7).
CGGC is level 2 fellow of CNPq and Investigador Correspondiente (CONICET).
This work was partially funded by Conacyt grant CB2009-134985.
\end{acks}

\bibliographystyle{spr-mp-sola}
\bibliography{libros}

\end{article}

\end{document}